\documentstyle[12pt]{article}
\def\J{$J/\psi$}
\def\j{J/\psi}
\def\X{$\chi$}
\def\x{\chi}
\def\P{$\psi'$}
\def\p{\psi'}

\def\e{\epsilon}

\def\be{\begin{equation}}
\def\ee{\end{equation}}

\def\lsim{\raise0.3ex\hbox{$<$\kern-0.75em\raise-1.1ex\hbox{$\sim$}}}
\def\gsim{\raise0.3ex\hbox{$>$\kern-0.75em\raise-1.1ex\hbox{$\sim$}}}

\def\NP{{ Nucl.\ Phys.\ }}
\def\PL{{ Phys.\ Lett.\ }}
\def\PR{{ Phys.\ Rev.\ }}

\def\PRL{{ Phys.\ Rev.\ Lett.\ }}

\def\ZP{{ Z.\ Phys.\ }}

\begin{document}

\centerline{\large{\bf Anomalous \J~Suppression}}

\medskip

\centerline{\large{\bf and the Nature of Deconfinement}}

\vskip 1cm

\centerline{D.\ Kharzeev, M.\ Nardi and H.\ Satz}

\bigskip

\centerline{Fakult\"at f\"ur Physik, Universit\"at Bielefeld}

\par

\centerline{D-33501 Bielefeld, Germany}

\vskip 1.5cm

\begin{abstract}

\par

We study the characteristic features of \J~suppression by deconfinement.
A first order deconfinement transition leads to an essentially
discontinuous onset of anomalous \J~suppression. The different energy 
densities required for the dissociation of different charmonium
states result in a two-step suppression pattern, in which first the \J's
from $\x_c$ decay are suppressed and subsequently the directly produced
\J states. Combining both features leads to a unique form of
\J~suppression in a deconfining medium.

\end{abstract}

\vskip 1.0cm

The suppression of \J~production observed in nuclear collisions, from
$p-A$ up to central $S-U$ interactions, can be understood as
pre-resonance charmonium absorption in normal nuclear matter
\cite{KS6,KLNS}. In contrast, recent data from $Pb-Pb$ show a
considerably stronger (and hence `anomalous') suppression \cite{NA50},
which may be a first indication \cite{Blaizot} of the predicted
signature for colour deconfinement \cite{Matsui}. The main difference
between such an interpretation and conventional attempts in terms of
hadronic comovers \cite{Gavin}-\cite{Cassing} lies in the fact that
deconfinement, as a critical phenomenon,
sets in at some specific temperature or density, while dissociation
in a hadronic medium must occur to some extent for all conditions.
The nature of the onset of anomalous suppression thus obtains a
particular qualitative importance in the search for its origin.

\medskip

A number of studies have tried to check whether the existing
experimental errors still allow a gradual change from slight
suppression in cooler $S-U$ to strong suppression in hotter $Pb-Pb$
collisions. Though such efforts quantitatively encounter considerable
difficulties, it would be more satisfactory to establish the onset of a
new state of matter already on a qualitative level. If the
deconfinement transition were of first order, the onset
of anomalous suppression would become discontinuous, which should remain
as a very abrupt onset even when averaged over an $E_T$ range.
Deconfinement, whether treated in terms of screening \cite{MTM,Karsch}
or in terms of gluon-charmonium interactions \cite{KS3}, always leads to
an easier dissociation for the excited states $\x_c(1P)$ and
$\p(2S)$ than for the ground state $\psi(1S)$. Since the \J's observed
in nuclear collisions are only about 60 \% directly produced $\psi(1S)$,
with the remainder coming from excited states, a deconfinement
suppression should show distinct different onsets \cite{Gupta}. Both the
abrupt onset of deconfinement and the two-step suppression pattern
by their very nature rule out all conventional scenarios. The aim
of this paper is therefore to study the effect of the transition
pattern on the observable suppression.

\medskip

Lattice studies of finite temperature QCD \cite{Engels,lattice} lead to
an equation of state of the form shown in Fig.\ 1, with a sharp
increase of the entropy density $s(T)$ in the vicinity of the
deconfinement temperature $T_c$, from a low hadronic value to a much
higher one in the quark-gluon plasma. In pure SU(3) gauge theory, the
transition is of first order; in full QCD, its order is
not yet clear at this time, but some recent work \cite{Kanaya} suggests also here
first order. In the present paper, we want to investigate the effect of
a first order transition in a nuclear collision environment, keeping in
mind in particular the small volume aspects of such reactions, and
study its consequences on \J~suppression.

\medskip

Consider the medium produced in the passage of two heavy nuclei.
It will be hadronic as long as the energy density remains below a
critical value $\e_c \equiv \e(T_c)$. For $\e > \e_c$, the hadronic
phase will become metastable and eventually undergo a transition to the
quark-gluon plasma. This transition proceeds through formation of
bubbles of the new deconfined phase, which must be large enough to
overcome the interface tension $\sigma$ between the two phases. Let us
determine this critical size \cite{Landau}. The free energy needed to produce a
bubble is
\be
F = - V~ \delta P + A ~\sigma , \label{1}
\ee
where $V$ is the volume of the bubble, $A$ its surface area, and
$\delta P = P_q - P_h$ the pressure difference between the deconfined
inside and the hadronic outside. The pressure difference can be
expressed in terms of the latent heat $L$ of the transition,
\be
\delta P = L~\tau, \label{2}
\ee
with $\tau\equiv (T-T_c)/T_c$ specifying the amount of overheating above
the critical temperature $T_c$. For spherical bubbles, Eq.\ (1) becomes
\be
F(r,T) = -{4 \pi r^3 \over 3} L ~\tau + 4\pi r^2~ \sigma. \label{3}
\ee
The derivative $dF/dr$ changes signs at the critical radius
\be
r_c(T) = {2\sigma \over L\tau}; \label{4}
\ee
for $r<r_c$, the bubble is unstable, shrinks and disappears again, while
for $r>r_c$ it expands and thus leads to the new phase.

\medskip
Precise results from finite temperature lattice QCD are at present available
only for pure SU(3) gauge theory \cite{SU(3)}; but these should provide
some indication of what it to be expected. Here one finds
\be
\sigma/T_c^3 = 0.0155 \pm 0.0016; ~~~L/T_c^4=1.80\pm0.18, \label{5}
\ee
which leads to
\be
r_c(T) \simeq {0.02 \over (T-T_c)} \label{6}
\ee
for the critical radius. An overheating by $T-T_c = 1$ MeV thus implies
a critical radius $r_c \simeq 4$ fm; the actual amount of overheating in
realistic high energy nuclear collision conditions is of course
difficult to estimate.

\medskip

Next, we have to determine the rate (per unit time) of bubble formation
in an overheated medium. It has the general form
\be
{dR \over dt} \sim \exp\{-F(r,T)/T\} \label{7},
\ee
which with Eq.\ (3) leads to
\be
{dR \over dt} \sim \exp\left(-{16 \pi \sigma^3 \over 3T L^2\tau^2}
\right). \label{8}
\ee
The evaluation of the pre-exponential factor requires detailed knowledge
of the dynamics of heating and the properties of the system \cite{Kapusta}.
The crucial
feature, however, is the non-analytic behaviour of expression (8) as the
critical temperature is approached, i.e., for $\tau \to 0$. It implies
that even a relatively small amount of overheating will bring the entire
system into the new phase, through the formation and expansion of bubbles
\cite{Miklos}.

\medskip

Let us now apply these general considerations to charmonium dissociation
in the produced environment. The essential aspect for this is that
the deconfinement transition proceeds through formation of bubbles of
some minimal size. While we obviously cannot perform a parameter-free
calculation, we want to show that the basic nature of the transition
leads to a characteristic onset of suppression which seems difficult to
obtain in any conventional approach.

\medskip

The minimal bubble size requires that the volume of hot and dense medium
produced in high energy collisions must exceed some critical value, so
that we need sufficiently central collisions of sufficiently heavy
nuclei. It moreover implies that the creation of the new phase occurs
abruptly in a significant fraction of the produced volume. If a specific
charmonium state is dissociated inside the bubble, this means a
discontinuous onset of suppression as function of centrality in a given
$A-A$ collision; the suppression sets in when the critical density is
overcome in a large enough volume. We shall here assume that the
given charmonium state is dissociated if and only if it finds
itsself within a bubble of deconfined phase.

\medskip

We thus consider the formation of `hot' bubbles of size $B$ consisting
of a deconfined medium. A charmonium state produced in a
nucleus-nucleus collision with impact para\-me\-ter $b$
and at point $s$ in the transverse plane of the medium will then have
the following survival probability $S(b,s)$:
\begin{itemize}
\item{}{If $(b,s)$ is not in the region $B$, or if $B<B_c$, where $B_c$
is a bubble of minimal size (6), or if the energy density in the
bubble, $\e(B)$, is less than that at the critical temperature,
$\e(B)<\e_c\equiv \e(T_c)$, then
\be
S_B(b,s) = 1,\label{9}
\ee
i.e., there is no (anomalous) suppression.}
\item{}{If however, $(b,s)$ is in $B$, the bubble volume is large
enough, $B\geq B_c$, and the energy density in the bubble is high
enough, $\e(B)\geq \e(T_c)$, then
\be
S_B(b,s) = 0, \label{10}
\ee
i.e., the charmonium state is destroyed.}
\end{itemize}
Let us now see what this means for the observed \J~suppression pattern.

\medskip

We have assumed in Eq.\ (10) that a given charmonium state is
dissociated
as soon as the medium within the bubble is deconfined. Studies of the
melting of quarkonia in hot matter \cite{MTM,Karsch} show that this is
correct for the $\x_c$~and the \P, but that the $1S$ state $\psi$
because of its much smaller size in fact needs a higher energy density
or temperature to be dissociated. The experimentally observed \J's are
about 32 \% due to \X~decay and about 8 \% to \P~decay \cite{decay};
the remaining 60 \% are directly produced as $\psi(1S)$. At the onset of
bubble formation, we thus expect an immediate onset of suppression for
that part of \J~production which comes from \X~and \P~decay. Directly
produced \J's will be suppressed only when the temperature of the
bubbles has increased sufficiently to reach the melting point of the
$\psi(1S)$; the above mentioned studies suggest that this takes place
at about 1.1 to 1.2 $T_c$. At this value, the experimental survival
probability should thus show a further decrease.
As a result, we obtain a characteristic two-step suppression pattern
due to the successive onsets of the different quarkonium states
\cite{MTM} - \cite{Gupta}.

\medskip

The crucial variables determining the critical behaviour of
\J~production are, as we have seen, the critical temperature or
energy density for deconfinement and the critical bubble size. In view of the
quantitative uncertainties in our picture, we will here fix both
empirically and then check whether the resulting values make sense.
A measure for the initial temperature of the produced medium is provided
by the density $n_w$ of participant or wounded nucleons, since the
number of wounded nucleons determines the number of secondary hadrons
and hence the energy density produced in the collision.

\medskip

To determine the actual survival probability of a charmonium state $i$
in a nuclear collision, we have to convolute its formation probability
$S^i_{Gl}(b,s)$ \cite{KLNS} with the bubble probability (9/10),
giving us
\be
S_i(b) = \int d^2s ~S^i_{Gl}(b,s)~S_B(b,s), \label{11}
\ee
where the role of the state $i$ enters through the value of the
temperature or energy density needed for melting. To relate this in
turn to the measured survival probability at constant transverse
hadron energy $E_T$, we have to convolute $S_i(b)$ with the $E_T-b$
correlation function $P(E_T,b)$ \cite{KLNS},
\be
S_i(E_T) = \int d^2b ~ P(E_T,b)~S_i(b) .\label{16}
\ee
Using the mentioned fractional composition of \J~production in terms of
\X, \P~ and direct $\psi$ states, we finally obtain
\be
S_{\j}(E_T) \simeq 0.4 ~S_{\x}(E_T) + 0.6~S_{\psi}(E_T), \label{17}
\ee
where we have combined the suppression of \J's produced through
intermediate \X~and \P~states in $S_{\x}$, neglecting any addditional
possible \P~suppression the hadronic environment outside of the bubbles.
To illustrate the resulting suppression pattern, we have to fix the
critical bubble size $B_c$, as well as the threshold values for deconfinement
and for the later onset of direct $\psi$ melting.

\medskip

In Fig.\ 2, we compare the \J~suppression data \cite{NA50} to the
pattern obtained from Eq.\ (17) with a transverse bubble radius 3.0 fm
and critical densities of wounded nucleons 
1.8 fm$^{-2}$ for the \X, 3.5 fm$^{-2}$ for the
direct $\psi$. With 1.5 pions of 0.5 GeV energy each produced 
per unit central rapidity for a wounded nucleon, this corresponds
to a deconfinement energy density of about 1.4 GeV/fm$^3$, in good
agreement with screened potential calculations \cite{MTM,Karsch} and 
estimates based on lattice QCD \cite{lattice}.
The energy density for the melting of the $\psi$ becomes about twice
that, again in accord with screened potential calculations
\cite{MTM,Karsch}. The location of the discontinuity at deconfinement
as well as the second drop at the onset of direct \J~suppression are
obviously dependent on the choice of the three parameters involved; we
have here not tried to optimize the description, but just chosen values
which reproduce approximately the behaviour of the published data. A
quantitative fit will be made as soon as the forthcoming higher
statistics data are available. The form of the suppression we have
obtained is, however, quite general. The abrupt onset and the
characteristic two-step nature of the suppression are direct
reflections of the transition order and of the successive
melting of the different charmonium components. The entire pattern is
qualitatively very different from any hadronic absorption picture. It
therefore would, if indeed observed in higher statistics data,
constitute a significant confirmation for the onset of deconfinement in
nuclear collisions.

\bigskip

\noindent{\large \bf Acknowledgement}

\bigskip

It is a pleasure to thank M.\ Creutz, O.\ Miyamura and 
K.\ Rummukainen for helpful discussions. D.\
K.\ and M.\ N.\ gratefully acknowledge financial support by the
GSI (Darmstadt) and the German Ministry for Education and Research
(BMBF).

\bigskip

\noindent{\large \bf Figure Captions}

\bigskip

\noindent Fig.\ 1: Temperature dependence of the entropy density in pure
SU(3) gauge theory \cite{Engels}

\bigskip

\noindent Fig.\ 2: \J~suppression for a first order deconfinement
transition and different dissociation points for \X~and direct $\psi$;
parameters are fixed using presently published data \cite{NA50}.


\newpage

\begin{thebibliography}{99}

\bigskip

\bibitem{KS6}D.\ Kharzeev and H.\ Satz, \PL B 366 (1996) 316.

\bibitem{KLNS}D. Kharzeev et al., \ZP C 74 (1997) 307.

\bibitem{NA50}M.\ Gonin [NA50], \NP A 610 (1996) 404c;\\
C.\ Louren{\c c}o [NA50], \NP A 610 (1996) 552c.

\bibitem{Blaizot}J.-P.\ Blaizot and J.-Y.\ Ollitrault, \PRL 77 (1996)
1703;\\
D.\ Kharzeev, \NP A 610 (1996) 418c.

\bibitem{Matsui}T.\ Matsui and H.\ Satz, \PL 178B (1986) 416.

\bibitem{Gavin}S.\ Gavin and R.\ Vogt, \PRL 78 (1997) 1006.

\bibitem{Capella}A.\ Capella et al., \PL B 393 (1997) 431.

\bibitem{Cassing}W. Cassing and C.-M.\ Ko, \PL B 396 (1997) 39.

\bibitem{MTM}F.\ Karsch et al., \ZP C 37 (1988) 617.

\bibitem{Karsch}F.\ Karsch and H.\ Satz, \ZP C 51 (1991) 209.

\bibitem{KS3}D.\ Kharzeev and H.\ Satz, \PL B 334 (1994) 155.

\bibitem{Gupta}S.\ Gupta and H.\ Satz, \PL B 283 (1992) 439.

\bibitem{Engels}G.\ Boyd et al., \NP  B 469 (1996) 419.

\bibitem{lattice}F.\ Karsch and E.\ Laermann, Rep. Prog. Phys. 56 (1993)
1347;\\ L.\ K\"arkk\"ainen et al., \NP B (Proc. Suppl.) 42 (1995) 460.

\bibitem{Kanaya}Y.\ Iwasaki et al., \PR D 54 (1996) 7010.

\bibitem{Landau}See e.g., L.\ D.\ Landau and E.\ M.\ Lifshitz, {\it Statistical Physics},
Pergamon Press, London 1958. 

\bibitem{SU(3)}K.\ Kajantie et al., \NP B 333 (1990) 100;\\
Y.\ Iwasaki et al., \PR D 49 (1994) 3540; \\
K.\ Rummukainen, ``Interface Tension with Lattice Monte Carlo",
Bielefeld Preprint BI-TP 96/59, 1996.

\bibitem{Kapusta}L.\ Csernai and J.\ Kapusta, \PR D 46 (1992) 1379.

\bibitem{Miklos}M.\ Gyulassy et al., \NP B 237 (1984) 477.

\bibitem{decay}Y.\ Lemoigne et al., \PL B 113 (1982) 509;\\
L.\ Antoniazzi et al., \PRL 70 (1993) 383.

\end{thebibliography}
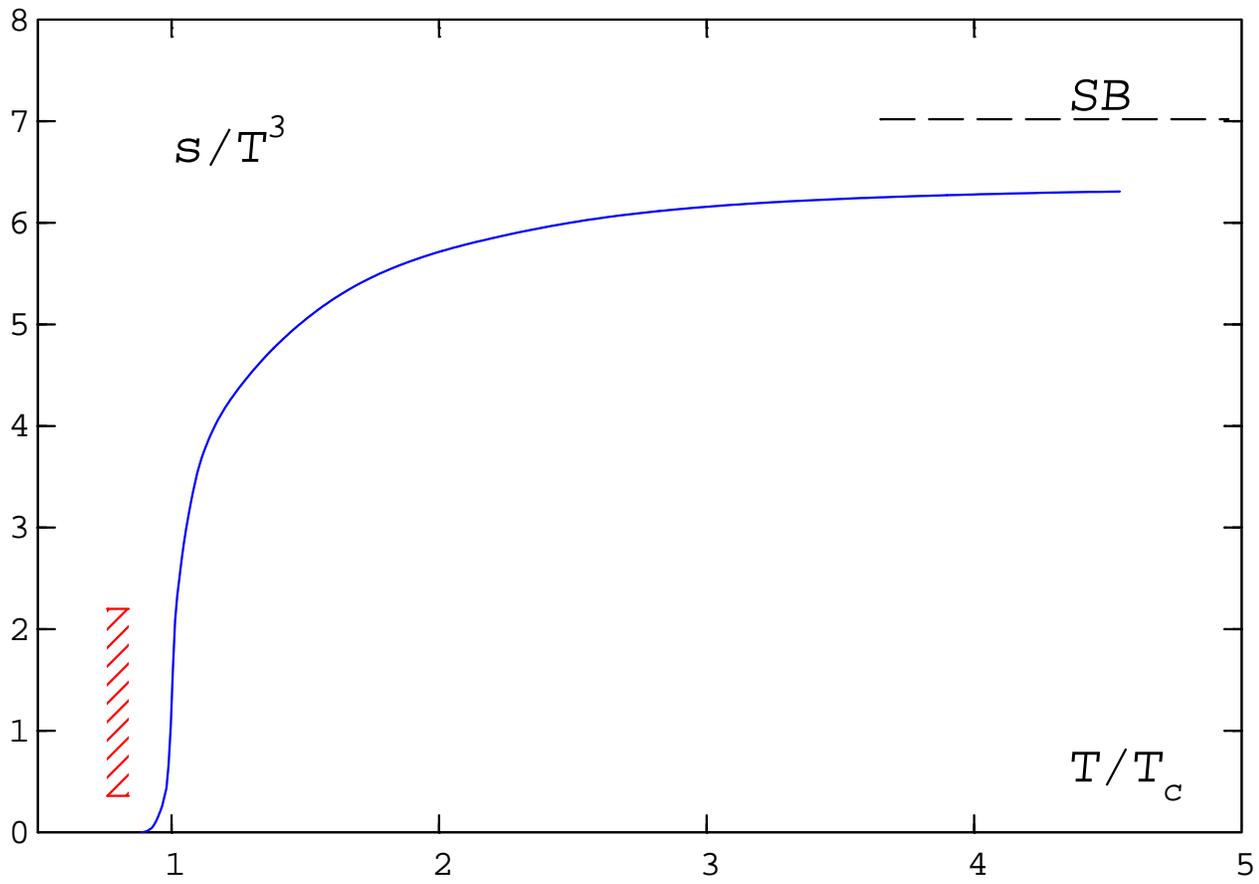
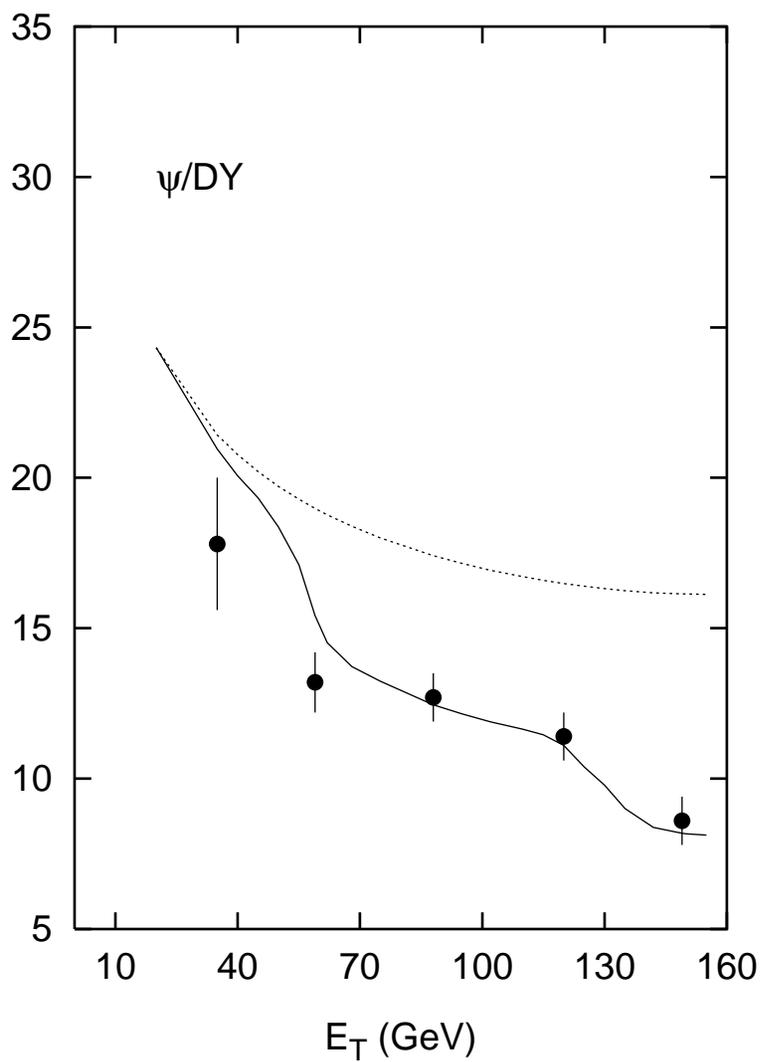
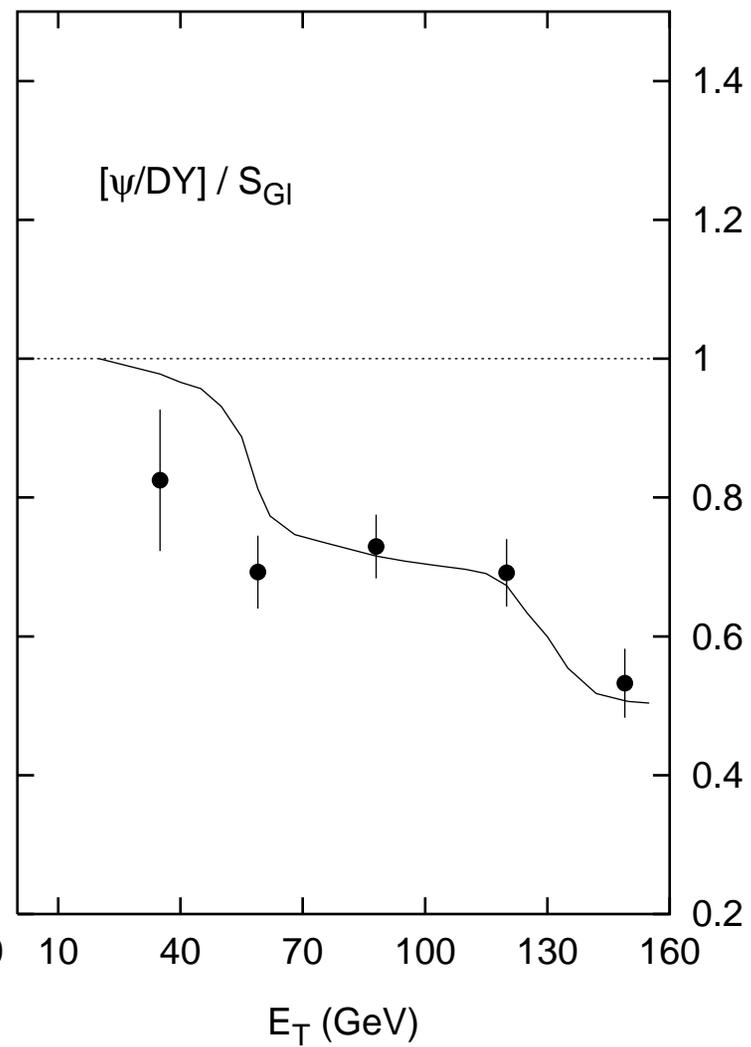
\end{document}